\documentclass[11pt]{revtex4}
\usepackage[utf8]{inputenc}
\usepackage{graphicx}
\usepackage{color}

\begin{document}

\title[Zero-mode mechanism for symmetry-breaking in von K\'arm\'an turbulence]{A zero-mode mechanism for spontaneous symmetry breaking in a turbulent von K\'arm\'an flow}

\author{B Saint-Michel, F Daviaud and B Dubrulle} 
\affiliation{Laboratoire SPHYNX, Service de Physique de l'\'Etat Condens\'e, DSM, IRAMIS, CEA Saclay, CNRS URA 2464, 91191 Gif-sur-Yvette, France}
\email{berengere.dubrulle@cea.fr}

\begin{abstract}
We suggest that the dynamical spontaneous symmetry breaking reported in a turbulent swirling flow at ${\rm Re}=40~000$ by Cortet {\itshape et al.}, Phys. Rev. Lett., 105, 214501 (2010) can be described  through a continuous one parameter family transformation (amounting to a phase shift) of steady states. We investigate a possible mechanism of emergence of such spontaneous symmetry breaking in a toy model of out-of-equilibrium systems. We show that the stationary states are solutions of a linear differential equation. For a specific value of the Reynolds number, they are subject to a spontaneous symmetry breaking through a zero-mode mechanism. The associated susceptibility diverges at the transition, in a way similar to what is observed in the experimental turbulent flow. Overall, the susceptibility of the toy model reproduces the features of the experimental results, meaning that the zero mode mechanism is a good candidate to explain the experimental symmetry breaking.
\end{abstract}

\maketitle

\section{Introduction}
Spontaneous symmetry breaking is a classical phenomenon in statistical and particle physics. From a macroscopic point of view, this loss of symmetry coincides with the loss of stability of the solution which respects the symmetries of the Hamiltonian (or evolution) operator of the problem and the emergence of a new set of stable solutions individually breaking the symmetry. Nevertheless, the set of solutions itself respects the broken symmetry to respect Curie's symmetry principle. Spontaneous symmetry breaking is also present in out-of-equilibrium systems, such as forced-dissipative flows. In the case where the dissipation is large, and the fluctuations very small, spontaneous symmetry breaking is well described through classical bifurcation theory~\cite{Berge1978,Busse1978,Cross1993} starting from linear or non-linear perturbations of the so-called ``basic state", the stationnary laminar solution of the Navier-Stokes equation at low Reynolds number~\cite{Manneville1995}.

When the fluctuations are much higher, and the symmetry breaking occurs over a turbulent background, however, tools are often missing to model the transition. This is the case for example when the symmetry breaking occurs for the mean state of a turbulent flow. This flow is stationary by construction, but differs from an usual basic state in the sense that it is solution of the ensemble time-averaged Navier-Stokes equation, differing from the plain
Navier-Stokes equation via a Reynolds stress tensor. This Reynolds stress represents the influence of all the degrees of freedom of the flow onto its average, and can, in general, only be computed via full solution of the NS equation. Therefore, the problem of instability of a mean turbulent flow cannot currently be tackled analytically or is too demanding numerically, unless a prescription (parametrization) of the Reynolds stress is provided. In the case of the plane Couette turbulent flow, for example, this was attempted by Tuckerman et al.~\cite{tuckerman_instability_2010} via the K - $\Omega$ closure model. 

In the present paper, we explore a new way to tackle the problem using tools inspired from statistical physics applied to a well-controlled laboratory model of spontaneous symmetry breaking, such as the von K\'arm\'an (VK) flow.  In this system, the flow is inertially forced by two counter-rotating impellers with blades, providing the necessary energy injection to set the system out-of-equilibrium. This energy is naturally dissipated through molecular viscosity, so that, for well controlled forcing protocols, statistically steady states can be established, that may be seen as the out-of-equilibrium counterpart of the equilibrium states of classical ideal systems~\cite{monchaux06, Monchaux2008}. Changing the forcing protocol for the VK flow leads to various transitions with associated symmetry breaking. In the sequel, we focus on the special case of parity symmetry breaking that has been reported in \cite{Cortet2010,Cortet2011}.
For exact counter-rotation (zero relative rotation)  of the impeller, the VK set up is exactly isomorphic to $O(2)$~\cite{nore2003}. Increasing the relative rotation between the two impellers, one induces symmetry breaking of the parity with respect to the vertical axis. Studying the flow response to this  symmetry breaking for a Reynolds number ranging from ${\rm Re} = 10^{2}$ (laminar regime) to ${\rm Re} \simeq 10^{6}$ (fully-developed turbulent flow), Cortet {\itshape et al.} observe a divergence of the flow susceptibility around a critical Reynolds number ${\rm Re}_c \approx 40\,000$. This divergence coincides  with intense fluctuations of the order parameter near ${\rm Re}_c$ corresponding to time-wandering of the flow between states which spontaneously and dynamically break the forcing symmetry. 

In this article, we investigate a possible mechanism of emergence of such spontaneous symmetry breaking in a toy model of an out-of-equilibrium system, derived from its equilibrium counterpart by adding forcing and dissipation. We show that the steady states of this model are subject to a spontaneous symmetry breaking through a zero-mode mechanism. We discuss how this model can be tuned to get qualitative agreement with the phase transition observed in the von K\'arm\'an experiment. We then show that the observed intense fluctuations of the order parameter near ${\rm Re}_c$ in the VK flow can  be described  through a continuous one parameter family transformation (amounting to a phase shift) of steady states that obey a Langevin equation.

\section{A toy model of steady states in a turbulent out-of equilibrium system}
Our goal is to build a simple model describing the steady states of an out-of-equilibrium system, that can the be mapped to an experimental VK turbulent flow. Our starting point is therefore the Navier-Stokes equations:
\begin{equation}
\frac{\partial {\bf u}}{\partial t} +{\bf u}\cdot \nabla {\bf u} =
   -\frac{1}{\rho}\nabla p + \nu \nabla^2 {\bf u}+ {\bf f},
\label{NS}
\end{equation}
where $\mathbf{u}$ is the solenoidal velocity field, $p$ the
pressure, $\rho$ the fluid density, $\nu$ its kinematic viscosity (that plays the role of control parameter in the sequel), and ${\bf f}$ a symbolic representation of the forcing whether it is described by a body force or through boundary conditions. In a steady state, where dissipation and forcing equilibrate on average, the total flow average energy $E =\int {\bf d^3x}~{\bf \overline{u^2}} /2$ ---~where $\overline{{\bf u}^2}$ denotes the time-average of ${\bf u}^2$~--- is conserved and satisfies:
\begin{eqnarray}
\partial_t E&=\int{\bf d^3x}\left(\overline{f_i u_j}-\nu \overline{\partial_j u_i \partial_j u_i}\right),\nonumber \\
&\equiv  {\cal F}+{\cal D}=0,
\label{stationaryE}
\end{eqnarray}

In some special situations, the Navier-Stokes equations have an interesting equilibrium counterpart, that is amenable to tools of classical statistical physics. This is the case when the flow is invariant by any rotation around a fixed axis, like in the VK flow we consider here. In such a case, the equilibrium counterpart is described by the axisymmetric Euler equations, the equilibrium of which can be derived using constrained extremalisation problems based on conservation laws~\cite{leprovost,naso09,thalabard13}. To describe the VK flow and obtain a toy model of its out-of-equilibrium steady states, we now adapt the constrained mean energy minimisation procedure of~\cite{naso09} to include forcing and dissipation through a procedure suggested by~\cite{werkley} in the framework of Jaynes maximum entropy principle. The equilibrium model of~\cite{naso09} can be described through only three fields~\cite{leprovost}: $u_\theta$, $\omega_\theta$ and $\phi$, where $u_\theta$ is the azimuthal velocity, $\omega_\theta$ is the azimuthal component of the vorticity, and $\phi$ is the streamfunction associated with the poloidal component of the velocity:
\begin{equation}
{\bf u} = u_\theta {\bf \hat{e}_\theta} + \nabla \times \left( \phi {\bf \hat{e}_\theta} \right) .\\
\label{defivit}
\end{equation}
The axisymmetry provides a simple relation between $\phi$ and $\omega_\theta$:
\begin{equation}
		 \Delta ( \phi ~\mathbf{\hat{e}_\theta}) = - \omega_\theta ~\mathbf{\hat{e}_\theta} \\
		 \label{eq:deltaphi}
\end{equation}
allowing the definition of a scalar operator $\mathcal{L}$. Equation~\ref{eq:deltaphi} can thus be expressed as:
\begin{equation}		 
		 \mathcal{L} \phi 						 = - \omega_\theta 
\end{equation}
This model is based on only three ideal invariants: the total energy 
$E = \langle \omega_\theta \phi+ u_\theta^2\rangle  / 2$, 
the total angular momentum 
$I =\langle r u_\theta\rangle$ 
and the helicity 
$H = \langle\omega_\theta u_\theta\rangle$, 
where $\langle \,.\, \rangle$ means spatial average. These integral constraints can be used to build a general Arnold energy-Casimir functional $A = E - \alpha I - \mu H$, the critical points of which provide the equilibria of the axisymmetric Euler equation~\cite{naso09}. In the out-of-equilibrium axisymmetric situation, the conservation of the energy necessitates balance of forcing and dissipation terms, i.e. ${\cal F}+{\cal D}=0$ (see Equation~\ref{stationaryE}), with ${\cal D}$ and ${\cal F}$
given by:
\begin{eqnarray}
{\cal D}&=\nu\int{\bf d^3x}\left(u_\theta\Delta u_\theta-\omega_\theta^2\right),\nonumber\\
{\cal F}&=\int{\bf d^3x}\left(f_\sigma u_\theta+f_\xi\omega_\theta\right).
\label{defiFetD}
\end{eqnarray}
To take into account this constraint, we follow~\cite{werkley} and introduce a Lagrange parameter $\zeta$ to build a new Arnold functional as: 
\begin{equation}
A^{\rm VK} = E -\alpha I -\mu H-\zeta\left({\cal D}+{\cal F}\right).
\end{equation}
 The critical points of this functional satisfy:
\begin{equation}
\left \{
\begin{array}{l l l}
	u_\theta 	  &=& B\phi+D(\nu)\omega_\theta+F \\
	\omega_\theta &=& Bu_\theta + Cr-D(\nu)\Delta u_\theta+G~,
\end{array}
\right .
\label{belhorsequilibre}
\end{equation}
where we have introduced for simplicity $B=1/\mu$, $C=-\alpha/\mu$, $D(\nu)=2\zeta\nu/\mu = D_0 \mu$ and $F=-D_0 f_\xi/$ et $G=-D_0 f_\sigma$.
This set of linearly coupled equations defines our toy model equation for describing the steady states of the out-of-equilibrium system. In that respect, $D$ is the control parameter caracterising the distance to equilibrium: for $D=0$, the steady states are the equilibria states of the axisymmetric Euler equation~\cite{naso09}. As $\vert D\vert$ is increased, the forcing and dissipation contribution grows and steady states deviate from the equilibrium solutions.

\section{Zero-mode analysis}

\subsection{Zero-mode mechanism in equilibrium}

In equilibrium systems, the spontaneous symmetry breaking occurs via a zero-mode mechanism that can be easily understood if we consider a general system for which  equilibrium is governed by a linear evolution operator $\mathcal{O}_\epsilon$:
\begin{equation}
 {\cal O}_\epsilon\phi=h~, 
\end{equation}
where $\epsilon$ is a control parameter, and $h$ represents an external driving field. Assuming for simplicity that $\mathcal{O}_\epsilon$ can be described by a matrix, with discrete spectrum, we can simply describe the solution of the previous equation as: 
\begin{equation}
 \phi(\epsilon)={\cal O}_\epsilon^{-1}h ~,
\end{equation}
if the kernel of ${\cal O}_\epsilon$ is empty. We then see that the equilibrium field follows the forcing symmetries and the susceptibility $\chi=\delta\phi/\delta h\vert_{h=0}$ is finite. In the case where the kernel of ${\cal O}_\epsilon$ is non-empty, the solution becomes: 
\begin{equation}
\phi(\epsilon)={\cal O}_\epsilon^{+}h+k~, 
\end{equation}
where $k$ is a zero-mode ---~element of the kernel~--- of $\mathcal{O}_\epsilon$ and $\mathcal{O}_\epsilon^+$ the pseudo-inverse of $\mathcal{O}_\epsilon$. In that case, the susceptibility diverges like $\chi\sim 1/\det(\mathcal{O}_\epsilon)$ and the field $\phi(\epsilon)$ follows the symmetry of the kernel of ${\cal O}_\epsilon$ as $h\to 0$. The range of values of $\epsilon$ for which the kernel of ${\cal O}_\epsilon$ is non-empty therefore corresponds to the situation with spontaneous symmetry-breaking solutions of ${\cal O}_\epsilon$ with diverging susceptibility. Despite being intrinsically an equilibrium result, we will see that this model still applies for von K\'arm\'an flows, where non-symmetric forcing conditions will play the role of the symmetry-breaking parameter $h$.

\subsection{The von K\'arm\'an case}

We now proceed to the zero-mode analysis of Equation~\ref{belhorsequilibre}. For this, we need to specify both the system geometry and the forcing. We consider a cylindrical geometry enclosed in the volume delimited above and below by surfaces $z=z_-$ and $z=z_+$, and radially by $0\le r \le R$. We will first consider the general case, for which $\phi = 0$ at $r = R$ but $\phi \neq 0$ at $z = z_\pm$. All velocity fields respecting the axisymmetry can be decomposed on the natural base $\phi_m$ of the Bessel-Fourier functions:
\begin{equation}
\label{e20}
\phi = \sum_{m=0}^\infty c_m \phi_m = \sum_{m=0}^\infty \mathcal{N}_m J_1\left(\frac{\lambda_m r}{R}\right)\left(a_m e^{iqz}+a_m^*e^{-iqz}\right)
\end{equation}
where the $c_m$ are complex coefficients and $\lambda_m$ is the $m^{th}$ zero of the $J_1$ function. The complex amplitude of the $m^{\rm th}$ mode is $a_m$, $\mathcal{N}_m$ is a normalisation constant and $q$ is the axial wavenumber. The $\phi_m$ also verify:
\begin{equation}
	\label{eq:def_K}
\mathcal L \phi_m = \left ( q^2 + \frac{\lambda_m^2}{R^2} \right ) \phi_m = K_m^2 \phi_m
\end{equation}
The set of $q$ will be specified by the forcing scenario: as we will see in the next section, the set for a body force scenario and for the boundary conditions scenario differ.

\subsection{Body force}
Without loss of generality, we can set $z_- = 0$, $z_+ = 2L$ and assume that $\phi=0$ at the boundaries. It is easy to verify that the $q$ selected by such conditions are of the form $q_n = n \pi / (2 L)$, $n$ being a positive integer. We therefore have to perform our projection on the $\phi_{mn}$ eigenfunctions: 
\begin{equation}
\label{eigenbody}
\phi_{mn}={\cal N}_{m}\, J_1\left (\frac{\lambda_{m}r}{R}\right )\sin\left (\frac{n\pi z}{2 L}\right ),
\end{equation}
with the normalisation constant
\begin{equation}
{\cal N}_{m} = \sqrt{\frac{2}{ J_2^2(\lambda_{m})}},
\end{equation}
These eigenfunctions are orthogonal with respect to the scalar product defined through the spatial average $\langle~~\rangle$: 
\begin{equation}
\left\langle f g \right\rangle \equiv \frac{1}{LR^2}\int_0^R \int_{0}^{2L} rdrdz~f(r,z) g(r,z).
\label{prod_sc1}
\end{equation}

The mode $(m,n)$ corresponds to $m$ cells in the $r$-direction and $n$ cells in the $z$-direction. We shall distinguish two kinds of modes, according to their properties regarding the symmetry ${\cal R}$ with respect to the plane $z=L$. The \emph{odd} (\emph{even}) eigenmodes are such that ${\cal R}\phi_{mn}=-\phi_{mn}$ (${\cal R}\phi_{mn}=\phi_{mn}$) and correspond to $n$ even (odd). We then proceed by decomposing all our fields onto the eigenfunctions through $(u_\theta,\omega_\theta,F,G)=\sum (s,x,f,g)_{mn}\phi_{mn}$. We can then recast Equation~\ref{belhorsequilibre} into $m\times n$ independent linear subsystems:
\begin{equation}
  \mathcal{M}_{mn} 
 	\left( \begin{array}{c}
 s_{mn}							  \\
 x_{mn}
 \end{array} \right ) = \left( \begin{array}{c}
 f_{mn}+ C \left\langle r\phi_{mn}\right\rangle\\
 g_{mn}
 \end{array} \right) 
 \label{eq:dynsyst}
\end{equation}
where
\begin{equation}
\mathcal{M}_{mn} = 
\left ( \begin{array}{cc} 
  1  			 &  -(D+BK_{mn}^{-2}) \\
-(B+DK_{mn}^{2}) & 1 				  
  \end{array} \right )
\end{equation}
and $K_{mn}^2 = (\lambda_m/R)^2 + q_n^2$. Furthermore, it may be checked that the corresponding solutions are actually minimising the functional $A^{\rm VK}$:
\begin{equation}
2 B A^{\rm VK} = \sum_{m,n} \Big( s_{mn} ~~ x_{mn} \Big) \bigg[ {\cal A}_{mn} \bigg ] \left (\begin{array}{c} s_{mn} \\ x_{mn} \end{array} \right ) + G s_n + F x_n + 2 C \langle r \phi_{mn} \rangle  .
\end{equation}
with
\begin{equation}
	{\cal A}_{mn} = 
	\left ( \begin{array}{cc}
		B + D K_{mn}^2 &	-1 \\
		-1		 & D + BK_{mn}^{-2} 
	\end{array} \right )	
\end{equation}
To ensure the existence of such minima, the matrix ${\cal A}_{mn}$ must verify ${\rm Tr} ({\cal A}_{mn}) \geq 0$ and ${\rm Det} ({\cal A}_{mn}) \geq 0$, yielding:
\begin{equation}
	\left \{
		\begin{array}{l}
			\left(1+ D B^{-1} K_{mn}^2\right)^2 \geq K_{mn}^2 B^{-2} \\
			\left(1+ D B^{-1} K_{mn}^2\right)  \left(  1+ K_{mn}^{-2} \right ) \geq  0 .
		\end{array}
	\right .	
\label{stable}
\end{equation}
Due to the symmetry properties of the modes, the $C$ parameter contribution vanishes for $n$ even.
The zero modes are obtained when the determinant of any $\mathcal{M}_{mn}$ subsystem is zero, occurring when:
\begin{equation}
\label{eq:kernel0}
D^2 K_{mn}^4+2BDK_{mn}^2+B^2-K_{mn}^2=0,
\end{equation}
i.e. for discrete set of values $D=D_{mn}$ 
\begin{equation}
\label{eq:DonneD}
D_{mn}^\pm=\frac{-B\pm \vert K_{mn}\vert}{K_{mn}^2}.
\end{equation}
The only stable solution is $D_{mn}^{{\rm sign}(B)}$ due to~Equation~\ref{stable}.
The corresponding zero-mode has the symmetry properties of the $mn$ Beltrami mode.

\subsection{Forcing through boundary conditions}
\label{forcing_bc}

We now consider that the flow is forced only through boundary conditions so that $F=G=C=0$.
Without loss of generality, we can assume $z_\pm=\pm L$ and that $\phi=0$ at the radial boundary and $\phi(\pm L,r)=\phi_\pm(r)$ at $z_\pm$. Eliminating $u_\theta$ and $\omega_\theta= -\mathcal{L} \phi$ in Equation~\ref{belhorsequilibre}, we obtain a single equation for $\phi$ as:
\begin{equation}
D^2 \mathcal{L}^2\phi+(1-2BD)\mathcal{L} \phi+ B^2\phi=0.
\label{newphi}
\end{equation}
This relation constrains the value of $q$, the axial wavenumber, through the values of $K$, determined by:
\begin{equation}
	\label{eq:dispersion_bf}
		D^2 K^4-(1-2BD)K^2+B^2	= 0
\end{equation}
and the existence of minima of $A^{\rm VK}$ in terms of the fields $\phi$, $u_\theta$ and $\omega_\theta$ is assured by the following conditions:
\begin{equation}
	\left \{
	\begin{array}{l}
			\left(1+DB^{-1}K^2\right)^2 \geq K^2 B^{-2},\\
			\left(1+DB^{-1}K^2\right) \left(1+K^{-2}\right) \geq 0,
	\end{array}
	\right .
	\label{condq}
\end{equation}
From Equation~\ref{eq:dispersion_bf}, we get the identity
\begin{equation}
	\label{eq:dispersion_id}
		(1+(D/B) K^2)^2	= (K/B)^2.
\end{equation}
Using Equation~\ref{condq} leads to the selection of the solution so that:
\begin{equation}
BD = \frac{\vert K /B\vert  -1}{(K/B)^2},
\label{DBK}
\end{equation}
that can be seen as the dispersion relation of the system.
Due to orthogonality of the eigenfunctions, the coefficients $a_m$ must satisfy the following 
properties to ensure that $\phi$ respects the boundary conditions:
\begin{eqnarray}
\label{boundarycond1}
 		\Re(a_m)\cos\left(qL\right)  &=+ \frac{1}{4{\cal N}_m^2}\left\langle J_1
 			\left(\frac{\lambda_m r}{R}\right) \left(\phi_+ + \phi_-\right)\right\rangle, \\
\label{boundarycond2}
		 \Im(a_m)\sin\left(qL\right) &=-\frac{1}{4{\cal N}_m^2}\left\langle J_1 					 	  			
		 \left(\frac{\lambda_m r}{R}\right) \left(\phi_+ - \phi_-\right)\right\rangle,
\end{eqnarray}
$\lambda_m$ being as before a zero of $J_1$.

\section{Response to an imposed symmetry breaking}
We now study the response to the system to a weak symmetry breaking, focusing on the VK-type $R_\pi$ symmetry breaking, obtained by breaking the symmetry with respect to the mid-plane $z_0=(z_-+z_+)/2$. These asymmetries, characterised by an amplitude $h \neq 0$, will be used to obtain detailed information on the spontaneous symmetry breaking process occurring in our perfectly symmetric model, as would an external magnetic field to understand the ferro-paramagnetic transition observed ---~for example, in the mean-field Ising model~--- at zero magnetic field. Both points of view (presence and absence of an external field) will be reconciled for vanishing asymmetries, or, in other terms, $h \to 0$.

In the non-equilibrium case, the $R_\pi$ symmetry of the system is achieved by choosing forcing or boundary conditions that are odd with respect to $z_0$. The weak symmetry breaking is then obtained by introducing a small even component in the forcing (that may include non-zero $C$ values) or boundary conditions depending on our scenario. In the equilibrium system, the $R_\pi$ symmetry is respected only for $C=0$; the weak-symmetry-breaking is obtained through a small increase of $C \propto h$, the amplitude of the symmetry breaking. In both equilibrium and non-equilibrium cases, we will consider the kinetic angular momentum $I$ of the flow: this simple quantity exhibits  as our symmetry-breaking order parameter due to its antisymmetry under $R_\pi$ symmetry.

\subsection{Body force}
When $F$ and $G$ are both odd, all the $(m,2n + 1)$ modes of $f$ and $g$ vanish. A ---~small~--- imposed symmetry breaking field $h$ will therefore be expressed as: $f_{m, 2n + 1} \propto h$, $g_{m, 2n + 1} \propto h$ for one (or more) couple of values of $m$ and $n$. As long as $D \neq D_{mn}$, the matrix $\mathcal{M}_{mn}$ is invertible and the solution fields $s$ and $x$ will be linear in $f$ and $g$, hence proportional to $h$. For vanishing symmetry breaking field, $I \to 0$ : the $R_\pi$ symmetry of the system is not spontaneously broken.
 
 At $D=D_{mn}$ the solution is a superposition of an odd function, the even mode, with amplitude $h$, and the Beltrami mode $\phi_{mn}$, with arbitrary amplitude. If this mode is odd ($n$ even), no spontaneous symmetry breaking occurs in the limit $h\to 0$. If this mode is even ($n$ odd), the solution has an even component even in the limit of $h=0$, and spontaneous symmetry breaking occurs, with non-zero value of the order parameter $I$, proportional to the amplitude of the eigenmode. The occurrence of this phenomenon depends on the value of $B$, that plays the role of a temperature, and on the value of $D$ that controls the distance to equilibrium. At equilibrium, $D=0$, and the spontaneous symmetry breaking can only occur for a discrete set of temperature $B=K_{m(2n+1)}$, as already noted in~\cite{naso09}. Out-of-equilibrium, $D \neq 0$ and the symmetry breaking can occur at any temperature.

\subsection{Forcing through boundary conditions}

\begin{figure}
	\centering
	\includegraphics[width=\textwidth]{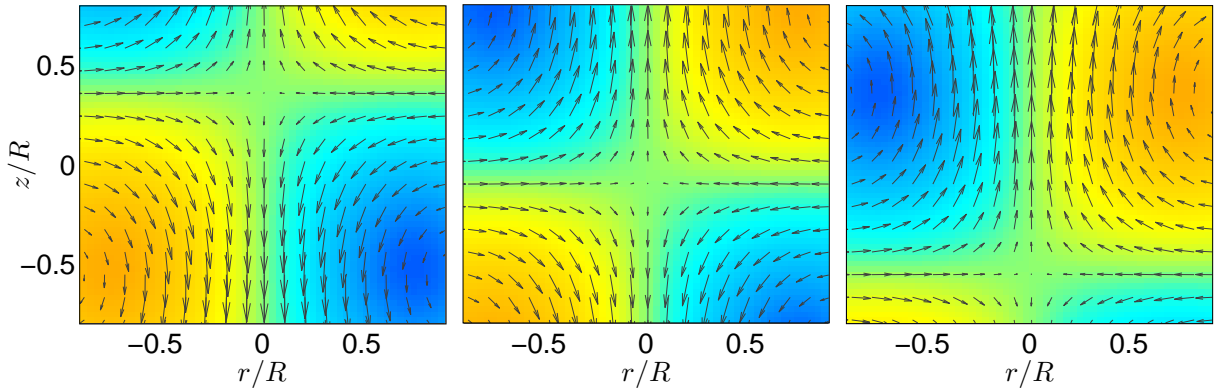}
	\caption{Theoretical symmetry breaking mode $(1,1)$ computed from equation~Equation~\ref{symmetrybreaking} for three different phases: a) $\psi_1\approx -\pi/4$ b) $\psi_1\approx 0$ c) $\psi\approx\pi/4$. The projection of this axisymmetric velocity field is displayed for two azimuth values ---~separated by $\pi$ radians~--- and correspond respectively to positive and ``negative" values of $r$.}
	\label{fig:goldstone}
\end{figure}

For a perfectly symmetric boundary forcing, we have to choose $\phi_\pm=\pm\phi_*$. The weak symmetry breaking can thus be parametrised by $\phi_+ + \phi_- \propto h$. In such a case, the boundary condition specified by Equations~\ref{boundarycond1} and~\ref{boundarycond2} leads necessarily to $\Re (a_m) \propto h$, except for the $q$ such that $\cos{qL}=0$, hence $q=q_n$, and $D = D_{m, 2n + 1}$ similarly to the previous section. So, for $D \neq D_{m(2n+1)}$, the solution is a superposition of $\sin$ (odd in $z$). At $D=D_{m(2n+1)}$, there is one coefficient $a_{m}$ with real part different from zero, with arbitrary amplitude. This provides, for $D = D_{m(2n+1)}$, a spontaneous symmetry breaking, associated with a diverging susceptibility when $D \to D_{m(2n+1)}$ (see Section~\ref{expsymbreak}). At $D=D_{m(2n+1)}$, and performing the spatial average, we see that $I\propto \Re (a_m)  R/L$, that can conveniently be written:

\begin{equation}
I\propto \frac{R}{L} \vert a_m\vert \sin (\psi_M),
\label{orderphase}
\end{equation}
where $\psi_m$ is the phase of the symmetry breaking eigenmode. In our system, the boundary conditions cannot specify simultaneously the amplitude and the phase of the symmetry breaking mode: for any given value of the parameter $D$ at the transition, there is a family of symmetry breaking modes $\phi_m$, labelled through a continuous parameter that we can choose as their phase $\psi_m$, so that:
\begin{equation}
\phi_m=J_1\left(\frac{\lambda_m r}{R}\right)\cos\left((2n+1)\pi z/2L+\psi_m\right).
\label{symmetrybreaking}
\end{equation}
An illustration of different members of this family corresponding to $M=1$, for different values of $\psi_m$ is provided in Figure~\ref{fig:goldstone}. When $\psi_1=0$, the solution is odd and does not break the system symmetry. As $\vert\phi_1\vert$ increases, the mode looses its parity symmetry, and $I$ becomes non zero. 

\subsection{Summary}
In both cases, the occurrence of the spontaneous symmetry breaking occurs when $D=D_{m(2n+1)}$, and is governed by the amplitude of the mode proportional to $\cos{(2n+1)z/2L}$, where $z$ spans from $-L$ to $L$. This breakdown can be parametrised by the phase of the symmetry breaking mode $\psi_m$ (in the case of forcing through boundary conditions), or its amplitude $a_m$ (in the case of the body forcing).

\begin{figure}
	\centering
	\includegraphics[width=0.75\textwidth]{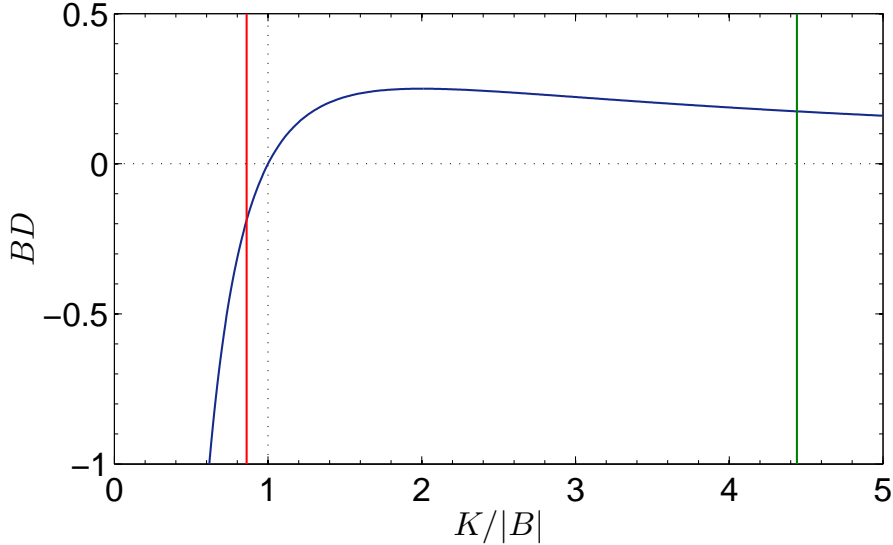}
	\caption{Dispersion relation and determination of the spontaneous symmetry breaking process. The vertical solid lines represent the  values of $K_{11}/B$ at $h=0.7$ and $R=1$, for two different values of $B$: (\textcolor{red}{---}) $B=5$, (\textcolor{green}{---}) $B=1$. The thick blue line displays $BD(K)$, given by Equation~\ref{DBK}. Their intersections yield the $D_{11}$ where zero modes appear and the first spontaneous symmetry breaking solution occurs.}
	\label{fig:solsre}
\end{figure}

The occurrence of the transition when the system is continuously driven out-of-equilibrium with $\vert D\vert$ increasing from $\vert D\vert=0$ depends on the value of $B$, that plays the role of a temperature. When $B\le K_{11}$, the first symmetry breaking transition occurs for negative values of $BD$, while when $B\ge B_{11}$ the first transition occurs for positive values of $BD$. This is illustrated in Figure~\ref{fig:solsre}. 

\section{Mapping to experiments}
\label{mapping}
\subsection{Calibration of the parameters}
\label{calibration}

We have derived our toy model of out-of-equilibrium starting from an equilibrium model of the Euler equation for an axisymmetric flow, the ideal limit of a force-free von K\'arm\'an flow with no dissipation. The equilibrium model was derived using well accepted principles of statistical mechanics. The toy model is just a convenient empirical generalization, based on Jaynes maximum entropy principle along the lines sketched in~\cite{werkley}. After suitable calibration of the parameters, it can nevertheless be mapped into the real experimental von K\'arm\'an flow to provide useful interpretation of the observations, as will now be demonstrated. 

\begin{figure}
	\centering
	\includegraphics[width=0.99\textwidth]{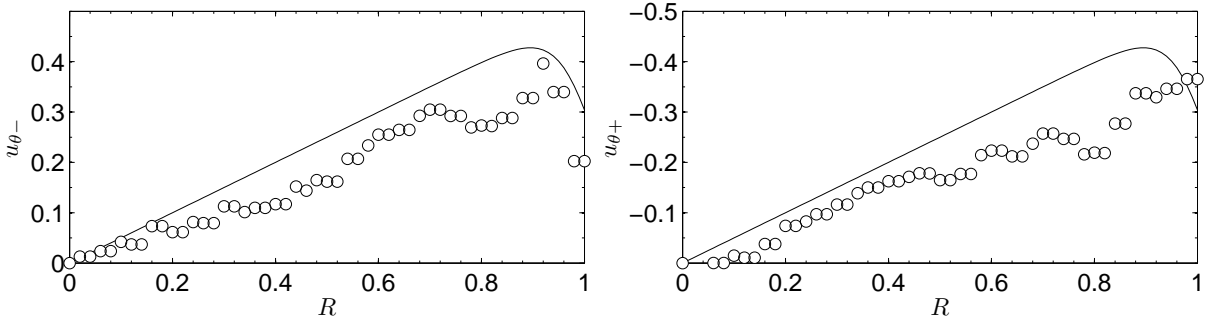}
	\caption{Comparison between the modeled (line) and experimental ($\circ$) azimuthal velocity boundary condition at $\theta=0, {\rm Re}=5145$. Left panel: at the bottom disk. Right panel: at the top disk.}
	\label{fig:compaboundy}
\end{figure}

The symmetry group of this experimental system is isomorphic to $O(2)$~\cite{nore2003} for an exact counter rotation of the impellers stirring the fluid. These impellers, of radius $R_{i}=0.975 R$, are located at $z=\pm L$ and rotating at a constant frequency $f$, following the scenario of section~\ref{forcing_bc}. The aspect ratio of the experiment is therefore $2L/R=1.4$. The control parameter is ${\rm Re}=2\pi R^2 f/ \nu$. A diverging susceptibility of the experiment to a small asymmetry $\theta$, associated with a phase transition, is achieved at ${\rm Re}_c \sim 40\,000$ or $90\,000$ depending on the selected order parameter~\cite{Cortet2010,Cortet2011}. No further transition is obtained as the Reynolds number is increased, until at least ${\rm Re}\sim 400\,000$. This suggests the following parameter calibration: the forcing is approximated by two antisymmetric boundary conditions for the azimuthal velocity at $z=\pm L$: $u_\pm=\pm U(r)$, with $U(r)$ given by
\begin{equation}
U(r)=0.5U_{*}\left(\tanh\left((r-r_*)/h_*\right)-1\right),
\label{modelrotation}
\end{equation}
where $U_{*}=E_{\rm ff}2\pi R f$, $h_*=0.08$, $r_*=1.1R_i$. For $\theta=0$, this shape models a smooth rotation of frequency $f$, over a size of the order of the disk radius $R_i$ and with efficiency $E_{\rm ff}$. This efficiency measures the maximum azimuthal velocity attained near the impeller, and was measured as  $E_{\rm ff}=0.5$ for the type of impellers considered here~\cite{RaveletPhD}. The quality of this modelling can be evaluated by comparison with 
the experimental azimuthal velocity profile at the disk location  obtained using the SPIV. This is done in Figure~\ref{fig:compaboundy}, for ${\rm Re}=5000$. We see that the agreement is reasonable ; however, we can see in these results that the forcing conditions are not perfectly symmetric. We have checked that this synthetic forcing remains compatible with the experimental data for any Reynolds number in the range ${\rm Re} \in [10^3,10^6]$, with the parameters set to the values described above. The boundary condition in $u_\theta$ can then be mapped into a boundary condition on $\phi$ using the mode decomposition of~Equation~\ref{e20} and the stationary condition of Equation~\ref{belhorsequilibre}, so that:
\begin{equation}
\phi_\pm=(B+DK^2)^{-1}u_\pm
\end{equation} 

 From the analysis of steady states of von K\'arm\'an flow at very large Reynolds number~\cite{monchaux06, Monchaux2008}, we get $B = - 4.5 < 0$. From the definition of $D$, we get: $D \propto 2 \zeta B/Re$. Assuming $\zeta \leq 0$, $BD$ is negative, $D$ is positive and the only transitions occurs for $D_{mn}$, given by Equation~\ref{eq:DonneD}. For $2L/R=1.4$, we have $K_{11}=4.44$, $K_{13}=7.75$, $K_{23}=7.37$, so that $B$ obeys $-\vert K_{12}\vert \le B\le -\vert K_{11}\vert$. Therefore, for any $K_{mn} \neq K_{11}$, $K_{mn}/\vert B\vert>1$, and the corresponding zero-modes are associated with positive values of $BD$ (see~Figure~\ref{fig:solsre}). Hence, this assumption on $\zeta$ implies that a single transition occurs for $D=D_{11}$, suggesting to map $D$ and ${\rm Re}$ through $D=D_{11}Re_c/{\rm Re}$. With explicit expressions both for $B$ and $D$, we can evaluate $K$ for any Reynolds number using Equation~\ref{eq:dispersion_bf} (or using Figure~\ref{fig:solsre} for a graphical determination of $K$).
 
\subsection{Comparison with experiments}
\subsubsection{Experimental phase transition}

\begin{figure}
	\centering
	\includegraphics[width = \textwidth]{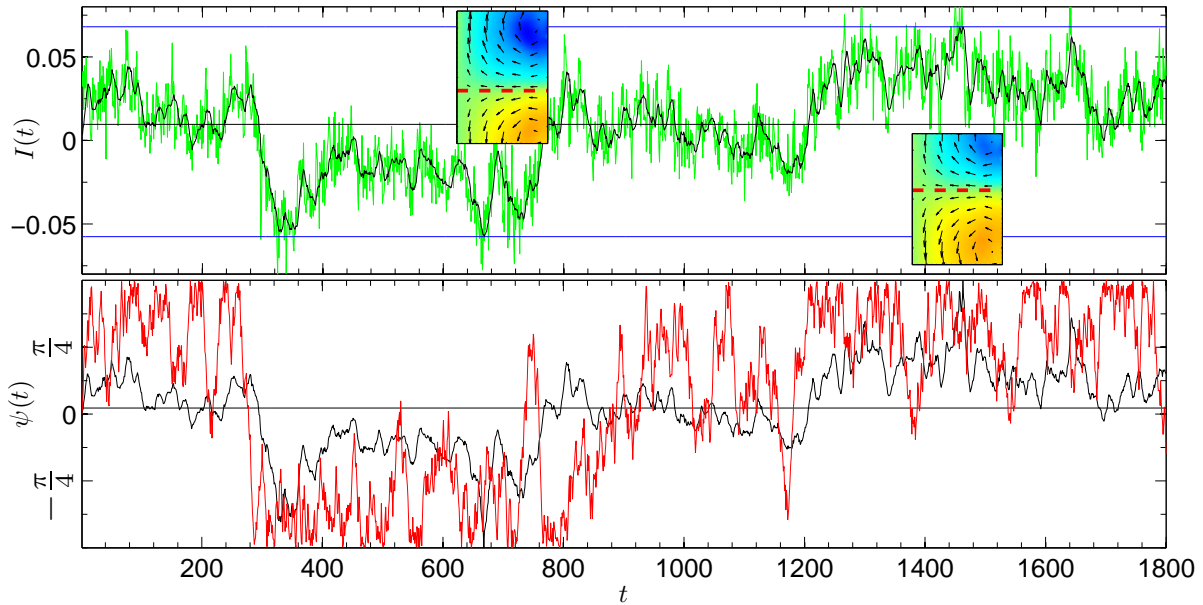}
	\caption{Top: Global angular momentum $I(t)$ as a function of time for an experiment performed at ${\rm Re}\sim 43,000$. Green lines are PIV data sampled at 15~Hz, and black lines correspond to 1~Hz low-pass filtered data. Eye-guide lines have been drawn: blue, at $I_0=\pm 0.04$; black, at $\overline{I(t)}$ to trace the time-average. The insets show the local-time average (over 50 frames) corresponding velocity field. On the bottom, we have displayed the symmetry breaking phase $\psi(t)$; black, computed from $I(t)$; red: computed from a Langevin model (see text for details). The horizontal black line traces the time average of $\psi$.}
\label{fig:IetPsi}
\end{figure}

In the experiment, the phase transition occurs at ${\rm Re}\sim 40\,000$~\cite{Cortet2011}. It is traced by large fluctuations in time of the order parameter $I(t)= \langle ru_\theta(t) \rangle$, where $u_\theta(t)$ is the instantaneous azimuthal velocity field, as measured through a stereoscopic particle image velocimetry. An example is provided in Figure~\ref{fig:IetPsi}, where one observes excursions of $I(t)$ away from $0$, till values that can reach $I=\pm I_0=\pm 0.04$, the amplitude of the symmetry breaking. During these excursions, the velocity field spontaneously breaks the symmetry, as illustrated in Figure~\ref{fig:IetPsi}, where the velocity field averaged over $50$ step times is shown: depending on the time around which the average is done, the velocity field presents a shear layer shifted downwards or upwards, 
and bears strong similarities with one of the member of the family breaking solutions $\phi_1$ illustrated in Figure~\ref{fig:goldstone}. From the phase measurements, one can compute the instantaneous phase of the symmetry breaking, through $\psi=\arcsin(I/I_0)$. This phase is shown in Figure~\ref{fig:IetPsi} and also displays large fluctuations away from zero.

\subsubsection{Velocity fields}
\begin{figure}
	\centering
	\includegraphics[width = 0.75\textwidth]{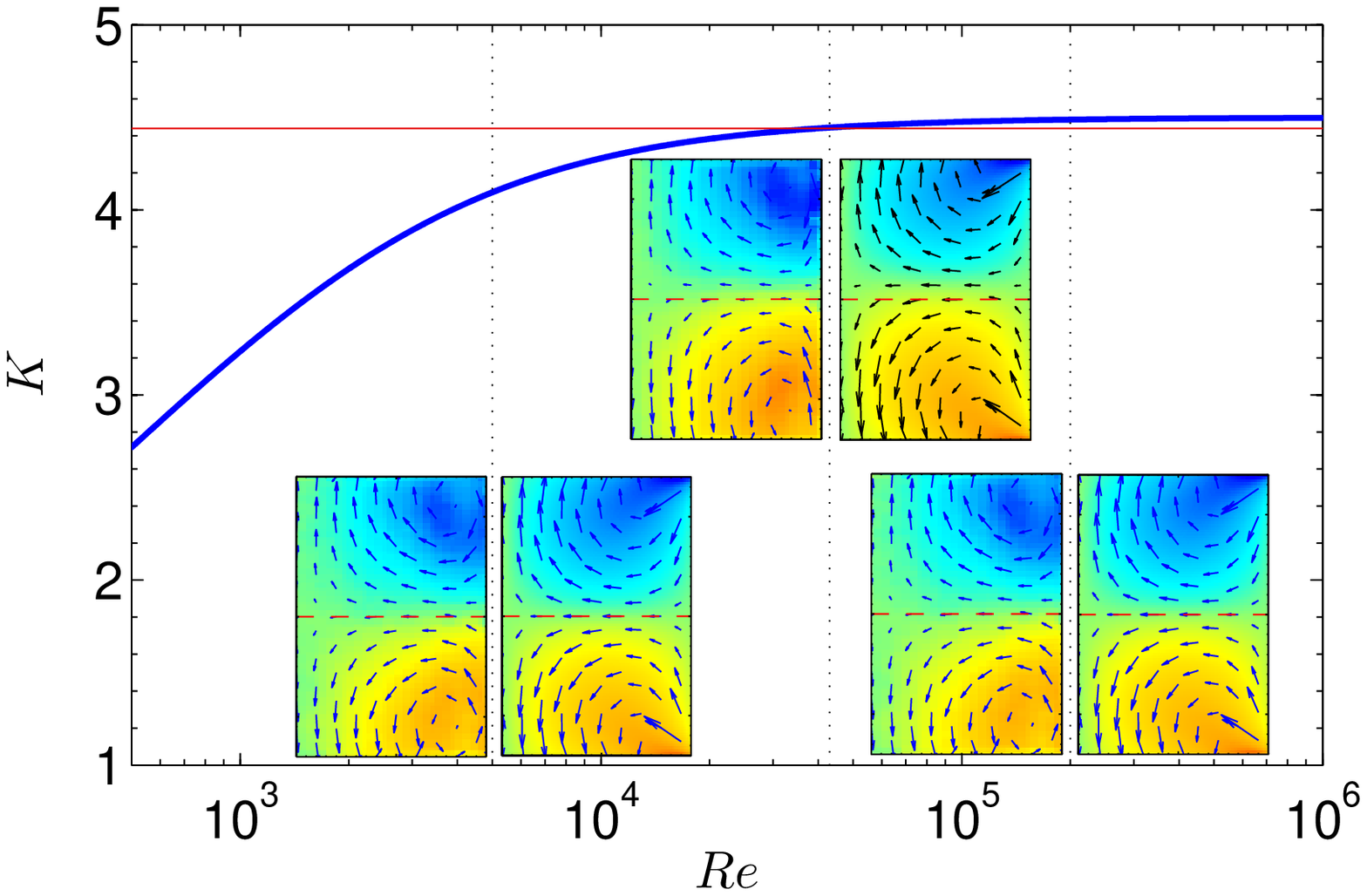}
	\caption{Test of the model for the turbulent von K\'arm\'an experiment: (\textcolor{blue}{---}) $K$ vs ${\rm Re}$ after the mapping of Section~\ref{mapping}. The red line indicates $K_{11}$. Its intersection with $K({\rm Re})$ occurs at  ${\rm Re}\approx40\,000$. The three couples of insets show the comparison between theoretical velocity fields (right), computed using the out-of-equilibrium toy model, and experimental velocity field(left) at ${\rm Re}=5000$, ${\rm Re}=43\,000$ and ${\rm Re}=200\,000$. The color codes the azimuthal field, the arrows indicate the $u_r,u_z$ field. The red dashed lines correspond to the $z = 0$ plane.}
	\label{fig:scenario1}
\end{figure}

A first test of the mapping can be obtained by comparing time-averaged (over 1500 frames) experimental velocity fields and theoretical velocity fields. These fields are calculated using the values of $B$ and $D$ of section~\ref{calibration} to obtain corresponding values of $K$ and $q$ (following equation~Equation~\ref{eq:def_K}). Finally, equations~Equation~\ref{boundarycond1} and~Equation~\ref{boundarycond2} define the projection coefficients $a_m$ which are used to reconstruct $\phi$ and $u$ using equation~Equation~\ref{e20} and the stationary condition of equation~Equation~\ref{belhorsequilibre}. 
This is done in Figure~\ref{fig:scenario1}, for different Reynolds numbers, one below the transition, one above the transition, and one at the transition. In the last case, the amplitude of the symmetry breaking mode has been chosen as $\Re (a_1) = I_0\sin(\overline{\psi})$, where $I_0$ and the time average $\overline{\psi}$ have been computed from Figure~\ref{fig:IetPsi}. One sees that the agreement is quite remarkable.

\begin{figure}
	\centering
	\includegraphics[width = 0.7\textwidth, trim = {5cm 14cm 5.5cm 3cm}, clip = true]{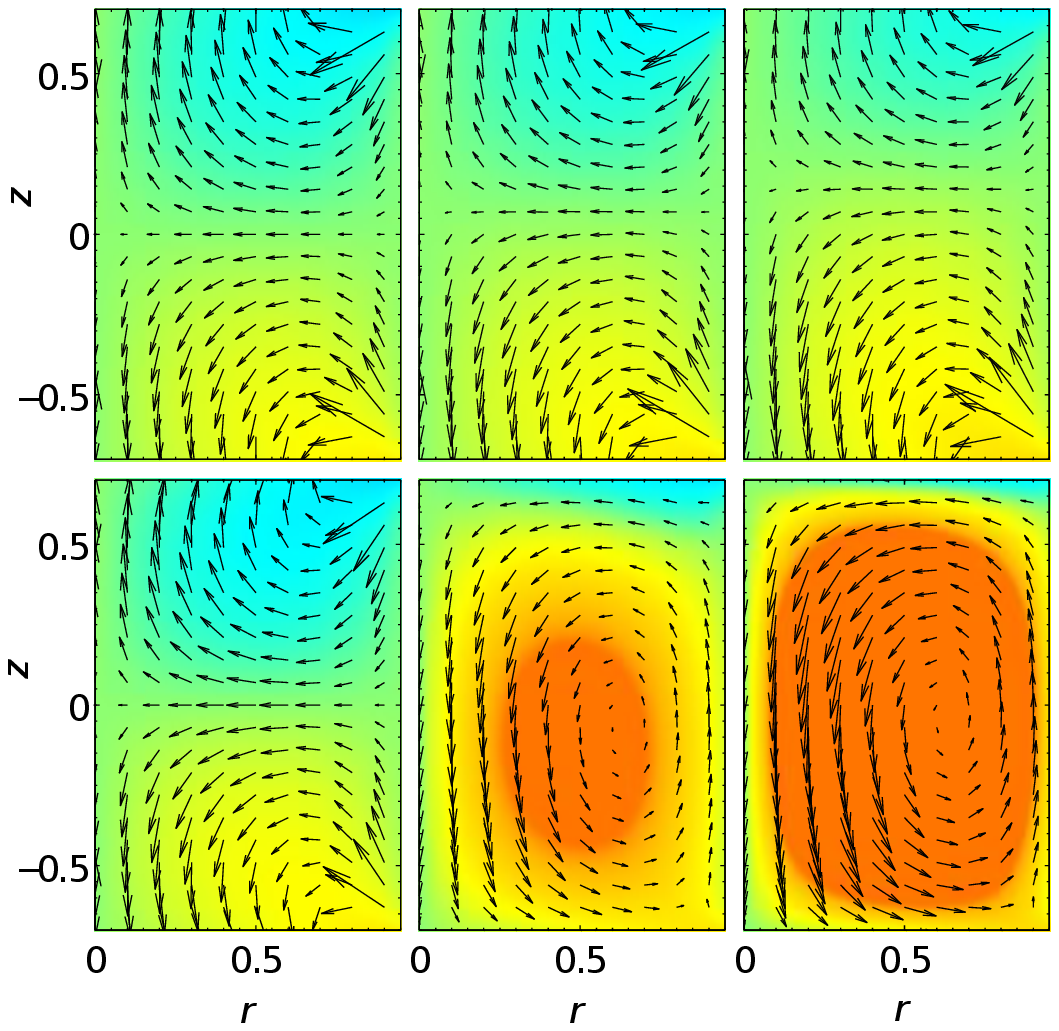}
	\caption{Velocity fields as computed from the toy model, at increasing $\theta$. The color codes the azimuthal field, the arrows indicate the $u_r,u_z$ field. Upper panel: at ${\rm Re}=3000$. Lower panel: at ${\rm Re}=30\,000$. From left to right: $\theta=0$, $\theta=0.1$ and $\theta=0.2$. The arrows for the last two panels have been rescaled with a factor $0.1$ for better readability.}
	\label{fig:velocitytheta}
\end{figure}

\subsubsection{Susceptibility to symmetry breaking}
\label{expsymbreak}
\begin{figure}
	\centering
	\includegraphics[width = 0.99 \textwidth]{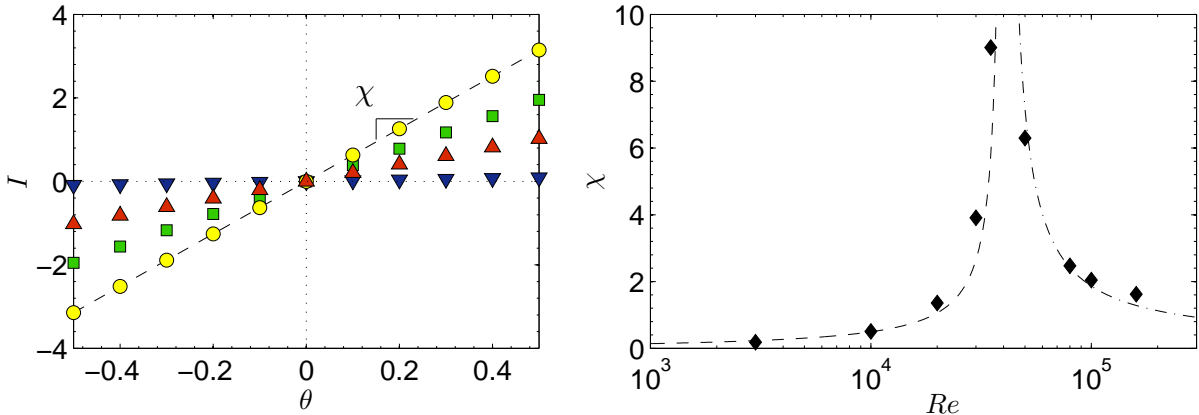}
	\caption{Left : $I$ as a function of $\theta$ in the toy model,  at ${\rm Re}=3000$ (triangles pointing downwards), ${\rm Re}=30\,000$ (squares), ${\rm Re}=50\,000$ (circles) and ${\rm Re}=80\,000$ (triangles pointing upwards). The color of the symbol codes $\log({\rm Re})$. Right panel : $\chi$ as a function of ${\rm Re}$ in the toy model. The lines are critical-like fits  $\chi=A_\pm 1/\vert T-T_c\vert$ where $T=1/\log({\rm Re})$ and $A_-=0.07$ (dashed line) and $A_+=0.14$ (dashed-dotted line).}
	\label{fig:SusceptideReToy}
	\label{fig:IdeThetaToy}
\end{figure}

A second test of the model is provided by the susceptibility to symmetry breaking. Computing the velocity fields as a function of $\theta$ for different Reynolds numbers using equations~(\ref{e20}) and~(\ref{boundarycond1}) with $f_\pm= f (1-\pm\theta)$, one indeed observes that the toy model response to symmetry breaking is quite different depending on whether the Reynolds number is far (e.g. ${\rm Re}=3000$) or close (e.g. ${\rm Re}=30\,000$) to its critical value. As can be seen in Figure~\ref{fig:velocitytheta}, the velocity field experiences a limited symmetry breaking at ${\rm Re}=3000$, with the central shear layer being progressively shifted towards the slowest impeller. In contrast, at ${\rm Re}=30\,000$, the velocity change with increasing $\theta$ is quite abrupt, resulting in an almost complete change towards a one-cell pattern (nearly antisymmetric velocity field) as soon as $\theta=0.1$. A quantitative estimate of this observation is provided by the variations of the quantity $I$ (computed from the model velocity fields) as a function of $\theta$. The results are presented in Figure~\ref{fig:IdeThetaToy}. One observes a linear increase of $I$ with $\theta$, with a slope depending on the Reynolds number: it is increasing until ${\rm Re}\sim 40\,000$ and decreasing after that value, in agreement with the observed behaviour in the experiment \cite{Cortet2011}. The resulting susceptibility to symmetry breaking, $\chi$, can then readily be obtained through a fit of $I(\theta)$, providing the result displayed in Figure~\ref{fig:SusceptideReToy}. One indeed observes a divergence of $\xi$ at ${\rm Re}={\rm Re}_c$, with a behaviour than can be fitted, like in the experiment, by a simple law: $\chi=A_\pm /\vert T-T_c\vert$, with $T=1/\log({\rm Re})$ and $A_\pm$ is a coefficient depending on whether the fit is performed before ($A_-$) or after ($A_+$) $T_c$. Here, we have used $A_-=0.007$ and $A_+=0.014$. Note that this corresponds to $A_+=2A_-$, a relation already observed in the experiment. This divergence observed in the experiment is about one magnitude larger than what is observed in the toy model, meaning that there is additional room for improving the model to make it best fit the experiment. 
However, the susceptibility of the toy model reproduces quite well the features of the experiment, meaning that the zero mode mechanism is a good candidate to explain the experimental symmetry breaking.

\subsubsection{Langevin model for the fluctuations of the kinetic momentum}
The observed spontaneous symmetry breaking fields recalled in Figure~\ref{fig:IetPsi} are very similar to one of the members of the family breaking solutions $\phi_1$. The intense fluctuations of the order parameter $I$, or equivalently of the phase of the symmetry breaking can therefore be viewed as a continuous time drift in between the different members of the symmetry breaking family, with a velocity given by $d\phi/dt$. From Figure~\ref{fig:IetPsi}, we observe that this velocity is quite fluctuating. 
This observation is in agreement with the fact that phase transition are always associated to fluctuations becoming very relevant, both in the zero mode directions and in the transversal direction. Focusing on the dynamics along the zero mode direction, we see that the simplest way to model it is through a Langevin equation, of the type:
\begin{equation}
\label{langevinmodel}
\left \{
	\begin{array}{rcl}
		\partial_t\phi				&=&v_G+\zeta \\
		\overline{\zeta(t)\zeta(t')}&=&Q\delta(t-t')
	\end{array}
\right .
\end{equation}
where $v_G$ is the phase velocity  and $\zeta$ is a delta-correlated noise of amplitude $Q$. For $v_G=0.7 f$ and $Q=3\pi$, the corresponding $\phi$ is shown in Figure~\ref{fig:IetPsi} and qualitatively matches the observed behaviour of the phase. 

In this framework, the phase transition in the experiment could therefore be interpreted as a zero mode symmetry breaking, with a mode that travels at a phase velocity $v_G$ with noisy disturbances caused by small scale velocity structures.

\section{Discussion}
In this paper, we have built a toy model of an out-of-equilibrium system, generalized from the corresponding 
equilibrium system to include forcing and dissipation following a suggestion by~\cite{werkley}. This procedure is based on Jayne's interpretation of statistical mechanics as a principle of insufficient reason. The corresponding  model displays spontaneous symmetry breaking, through a zero mode mechanism. We have shown that the model can be simply mapped to a real von K\'arm\'an experiment, by calibrating 3 parameters: the efficiency of the impeller $E_{\rm ff}$, the temperature $B$ and the critical Reynolds number ${\rm Re}_c$. Once these 3 parameters have been specified, all features of the experimental spontaneous symmetry breaking can be reproduced in a quantitative way and the time dynamics of the fluctuations of the order parameter can be interpreted through a  Langevin equation. 
The detail of the model can also be adapted to reproduce the observations of de la Torre and Burguete~\cite{Burguete2007} for example by inserting a three-well potential in the first line of Equation~\ref{langevinmodel} like in~\cite{Burguete2009}.

From the point of view of statistical mechanics, this provides interesting open questions regarding the modelling of out-of-equilibrium system: is the empirical procedure suggested by~\cite{werkley} to go from equilibrium to out-of-equilibrium correct? In the specific case of axisymmetric flows, we note that it leads to a regularisation of the equilibrium model by including bounds on the phase space, which would otherwise be infinite~\cite{thalabard13}, and on the vorticity fluctuations that are diverging in the equilibrium model. Other ways to limit the phase space are possible, for example by considering Casimir invariants of higher degrees~\cite{thalabard13}. It would be interesting to compare the corresponding regularized equilibrium models with experiment, to test whether the toy model we built here is the optimal one, and whether the agreement is purely fortuitous.

From the point of view of turbulence, the open question is how to compute the three calibrated parameters from general principles. Another question concerns the generality of the zero mode mechanism evidenced here. Does this mechanism work in other turbulent systems, like the stripe formation in the plane Couette flow \cite{PCF1, PCF2, PCF3, PCF4}, the mean flow reversals in  rotating Rayleigh-Bénard convection~\cite{RotConv1,RotConv2} or even  spontaneous transitions observed in natural systems (zonal to blocked pattern transition in northern hemisphere winds~\cite{BlockZon} patterns, Kuroshio currents~\cite{Kawabe1995})?

Finally, we observe that the  fluctuations of the order parameter near ${\rm Re}_c$ in the VK flow can  be described  through  continuous phase shifts of steady states. The corresponding set of solutions becomes then invariant under vertical translational symmetry. Such symmetry occurs only in an ideal system, composed of an infinite cylinder with neither forcing nor vertical boundary conditions. We are thus faced with a case where the {\sl turbulent} 
system undergoes a bifurcation that statistically restores the symmetry of the ideal, unforced system at large scale. This is then  a variant  of the H1 symmetry restoration hypothesis of \cite{frischbook}, albeit for the large scales.

\section*{Acknowledgements}
We thank the CNRS and the CEA for support, C. Herbert for interesting discussions and P.-P. Cortet and E. Herbert for sharing the experimental data with us.

\rule{0cm}{0.5cm}

\bibliographystyle{unsrt}
\bibliography{zmodes}

\end{document}